\documentclass[journal]{IEEEtran}
\ifCLASSINFOpdf
\else
\fi

\usepackage{xcolor}
\usepackage{microtype}
\usepackage{subfigure}
\usepackage{booktabs} 
\usepackage{amssymb}
\usepackage{multirow}
\usepackage{mathtools}
\usepackage{bm}
\usepackage{esvect}
\usepackage[square,compress,sort,comma,numbers]{natbib}

\begin{document}
%
\title{Robust End-to-End Focal Liver Lesion Detection using Unregistered Multiphase Computed Tomography Images}
%
%
%

\author{Sang-gil~Lee*,~Eunji~Kim*,~Jae~Seok~Bae*,~Jung~Hoon~Kim,~and~Sungroh~Yoon,~\IEEEmembership{Senior Member, IEEE\\
*Equal contribution}
\IEEEcompsocitemizethanks{
\IEEEcompsocthanksitem J.~Kim and S.~Yoon are corresponding authors.\protect\\
E-mail: jhkim2008@gmail.com; sryoon@snu.ac.kr
\IEEEcompsocthanksitem S.~Lee and E.~Kim are with the Department of Electrical and Computer Engineering, Seoul National University, Seoul 08826, Korea.
\IEEEcompsocthanksitem J.~Bae is with the Department of Radiology, Seoul National University College of Medicine, Seoul 03080, Korea, and also with the Department of Radiology, Seoul National University Hospital, Seoul 03080, Korea.
\IEEEcompsocthanksitem J.~Kim is with the Department of Radiology, Seoul National University College of Medicine, Seoul 03080, Korea, and also with Institute of Radiation Medicine, Seoul National University Medical Research Center, Seoul 03080, Korea.
\IEEEcompsocthanksitem S.~Yoon is with the Department of Electrical and Computer Engineering, Seoul National University, Seoul 08826, Korea, and also with Interdisciplinary Program in Artificial Intelligence, Seoul National University, Seoul 08826, Korea.
}
\thanks{This work was supported by the National Research Foundation of Korea (NRF) grant funded by the Korea government (Ministry of Science and ICT) [2018R1A2B3001628], the Interdisciplinary Research Initiatives Program from College of Engineering and College of Medicine, Seoul National University [800-20170166], Institute of Information \& communications Technology Planning \& Evaluation (IITP) grant funded by the Korea government(MSIT) [NO.2021-0-01343, Artificial Intelligence Graduate School Program (Seoul National University)], and the BK21 FOUR program of the Education and Research Program for Future ICT Pioneers, Seoul National University in 2021.}
\thanks{Manuscript received April 1, 2021; revised October 11, 2021; accepted November 11, 2021.}
\thanks{© 2021 IEEE.  Personal use of this material is permitted.  Permission from IEEE must be obtained for all other uses, in any current or future media, including reprinting/republishing this material for advertising or promotional purposes, creating new collective works, for resale or redistribution to servers or lists, or reuse of any copyrighted component of this work in other works.}
\thanks{Digital Object Identifier 10.1109/TETCI.2021.3132382}
}

%
%

\markboth{IEEE TRANSACTIONS ON EMERGING TOPICS IN COMPUTATIONAL INTELLIGENCE}%
{Lee \MakeLowercase{\textit{et al.}}: Robust End-to-End Focal Liver Lesion Detection using Unregistered Multiphase Computed Tomography Images}
%



\maketitle

\begin{abstract}
The computer-aided diagnosis of focal liver lesions (FLLs) can help improve workflow and enable correct diagnoses; FLL detection is the first step in such a computer-aided diagnosis. Despite the recent success of deep-learning-based approaches in detecting FLLs, current methods are not sufficiently robust for assessing misaligned multiphase data. By introducing an attention-guided multiphase alignment in feature space, this study presents a fully automated, end-to-end learning framework for detecting FLLs from multiphase computed tomography (CT) images. Our method is robust to misaligned multiphase images owing to its complete learning-based approach, which reduces the sensitivity of the model’s performance to the quality of registration and enables a standalone deployment of the model in clinical practice. Evaluation on a large-scale dataset with 280 patients confirmed that our method outperformed previous state-of-the-art methods and significantly reduced the performance degradation for detecting FLLs using misaligned multiphase CT images. The robustness of the proposed method can enhance the clinical adoption of the deep-learning-based computer-aided detection system.
\end{abstract}


%
\IEEEpeerreviewmaketitle

\section{Introduction}
\IEEEPARstart{H}{epatocellular} carcinoma (HCC) is the fifth most common cancer type and the second leading cause of cancer-related deaths worldwide \cite{mittal2013epidemiology}. Compared with other malignancies, HCC is unique in that it can be diagnosed based on imaging without histologic confirmation in patients with a high risk of developing HCC. Noninvasive diagnostic criteria established by major guidelines include hyperenhancement in the arterial phase and washout in the portal or delayed phase images \cite{european2012easl, omata2017asia, heimbach2018aasld}; these are considered to be associated with hemodynamic changes that occur during hepatocarcinogenesis \cite{kitao2009hepatocarcinogenesis}. Therefore, it is essential to obtain and interpret multiphase images of dynamic contrast-enhanced computed tomography (CT) or magnetic resonance images after the injection of contrast media to diagnose HCC.

When a radiologist diagnoses a focal liver lesion (FLL) in real-world clinical practice, he or she performs a comprehensive assessment of the lesion by characterizing its dynamic enhancement pattern on multiphase images. First, the precontrast, arterial, portal, and delayed phase images are aligned at a level where the lesion is best represented. Thereafter, the enhancements of the lesion, surrounding liver, and adjacent vascular structures on multiphase images are compared. If the lesion demonstrates a radiological hallmark of HCC (i.e., arterial enhancement and portal or delayed washout) and it is discovered in a patient with chronic liver disease, then a noninvasive diagnosis of HCC can be performed \cite{choi2014ct}.

This complex process of detection and characterization using multiphase images differs from the traditional application of deep learning in detecting FLL using only single-phase images \cite{bilic2019liver,li2015automatic, ben2016fully, christ2016automatic, christ2017automatic, han2017automatic, li2018h, zlocha2019improving}. The evaluation of multiphase images is not merely the detection of a lesion, i.e., the alignment of images in multiple phases is required. If multiphase images are misaligned because of motion artifacts or respiratory movements, the diagnostic performance of deep learning would be undermined.
In the detection task, wherein the model should accurately locate the lesion from images, the performance is significantly affected by misregistration because the model attempts to predict the bounding box based on an incorrectly aligned feature grid. Therefore, it is critical that the model accurately addresses the misalignments in multiphase images.

Existing literature regarding deep-learning-based systems presents challenges in processing multiphase images using external registration methods, and the deep-learning-based model is assumed to operate on a perfectly aligned image space \cite{lee2018liver, liang2018combining, liang2019multi, hasegawa2020automatic}. Although this two-step approach renders the task easier to the deep-learning-based model, its implementation in real-world clinical practice presents several disadvantages: the diagnostic pipeline becomes complicated and the selection of registration method becomes a confounding factor for the model’s performance. Achieving multiphase registration using external tools often requires time-consuming optimization processes and domain-specific expertise, rendering the deployment of such diagnostic models to the real-world clinical environment challenging.

This study presents a fully-automated, end-to-end deep-learning-based computer-aided detection system that detects FLLs from multiphase CT images, leveraging a high-performance single-stage object detection model. The proposed model achieves a significantly higher performance compared to previous methods when detecting FLLs. More importantly, it provides robustness to misalignment, which has yet to be addressed in the literature. This is achieved by our novel trainable method, i.e., \textit{attention-guided multiphase alignment} in a deep feature space. The model can leverage more extensive information present in multiphase CT images through a careful injection of parameterized task-specific inductive bias directly into neural network architectures. The method combines several neural subnetworks from distinct design disciplines to enable an inductive functionality required for the task. A grouped convolution \cite{krizhevsky2012imagenet} decouples the feature extraction of multiphase images. A self-attention mechanism \cite{vaswani2017attention} localizes and selects the salient aspect for detection. A deformable convolution kernel \cite{dai2017deformable, zhu2019deformable} compensates for the spatial and morphological variances inherent to multiphase CT.

We provide an extensive and ablative analysis of the proposed method used in single-stage object detection models \cite{liu2016ssd, deng2018pixellink}, using a custom-designed large-scale dataset containing multiphase CT images of 280 patients with HCC. Experimental results demonstrate that the method significantly outperforms previous state-of-the-art models for detecting FLLs under various conditions of registration quality.
\section{Related Studies}

\subsection{Computer-aided Liver Lesion Diagnosis}
Several deep-learning-based segmentation and detection approaches have been developed to detect FLLs based on portal phase CT images \cite{li2015automatic, ben2016fully, christ2016automatic, christ2017automatic, han2017automatic, li2018h, zlocha2019improving}. \cite{zlocha2019improving} presented a system developed using a state-of-the-art object detector (RetinaNet \cite{lin2017focal}), introducing an anchor optimization method and utilizing dense segmentation masks approximated from weaker labels called Response Evaluation Criteria in Solid Tumors (RECIST). However, their system was evaluated on the single-phase DeepLesion \cite{yan2018deeplesion} dataset, which does not challenge the accurate handling of multiphase data.

Some studies proposed an end-to-end detection of FLLs with multiphase CT using grouped convolutional layers and recurrent neural networks for inter-phase feature extraction~\cite{lee2018liver,liang2018combining,liang2019multi}. In \cite{lee2018liver}, a grouped single-shot multibox detector (GSSD) is proposed using an explicit decoupling feature extraction with a grouped convolution~\cite{krizhevsky2012imagenet}. This method has been successfully adapted to different neural architectures~\cite{liang2019multi}. However, it does not address the misalignment between modalities from the perspective of the model architecture. The existing approaches~\cite{lee2018liver,liang2018combining,liang2019multi} rely on an existing registration technique, that is, a preprocessing~\cite{dong2015non}, and assume a perfect intermodal alignment. Hence, they do not provide robustness to misalignments, and the performance of the model cannot be guaranteed for data with an unknown registration quality.
To our best knowledge, this study is the first to present an end-to-end, deep-learning-based detection of FLLs from a large-scale multiphase, unregistered CT database.

\subsection{Attention Mechanism}
The concept of attention has recently become a cornerstone in deep learning research~\cite{hassabis2017neuroscience}, and the mechanism has been integrated into a wide range of tasks~\cite{sutskever2014sequence, brock2018large, shen2018natural}.
In terms of medical images, the attention mechanism provides a trainable module that dynamically selects salient features present in an image that are effective for specified tasks~\cite{schlemper2019attention,zhang2019attention,sinha2020multi,wang2019abdominal}. A soft-attention gating module proposed in \cite{schlemper2019attention} constructs an attentive region in the current feature using the feature from a different scale, which is integrated to Sononet~\cite{baumgartner2016real} for image classification and U-net~\cite{ronneberger2015u} for image segmentation~\cite{oktay2018attention}.

The previous studies in medical images have used the attention mechanism to ensure that the network focuses on the target area (e.g., lesion, viscera) in a single-phase image. In contrast, this study utilizes the attention mechanism to exploit the salient features from multiphase images and induce the model to guide the learning-based interphase alignment and cross-modal feature aggregation for the detection task.

\subsection{Multimodal Registration}

Medical image registration is crucial to achieving an accurate and robust computer-aided diagnosis system. Various approaches have been proposed, including statistical \cite{ashburner2007fast, glocker2008dense, dalca2016patch} and learning-based ones \cite{yeo2010learning, balakrishnan2019voxelmorph, kim2019unsupervised, lee2019image}, which aim to find an optimal transformation between a pair of images. By contrast, the multiphase registration task that we posed to the model is \textit{implicit} in the deep feature space, not in the image space. This study does not directly provide the model with either the reference phase information or the registration-focused loss function.
The implicit task is inherent in the dataset, wherein the model is required to dynamically focus on the misaligned information and adjust the offsets of the deep feature map to achieve the maximum object detection performance.
\section{Multiphase Data}

\subsection{3D CT Data Collection}
We searched the image and pathology database of Seoul National University Hospital and retrospectively gathered data of patients who had undergone liver protocol CT before surgery for suspected HCC and were pathologically confirmed to have HCC based on surgical specimens. A board-certified radiologist reviewed the CT images of all the patients and included patients who had single HCCs with the typical enhancement pattern of HCC (i.e., arterial phase enhancement and portal/delayed phase washout). Finally, 280 patients (233 men; mean age $\pm$ standard deviation, 58.3 $\pm$ 10.4 years) with single HCCs were determined (mean size $\pm$ standard deviation, 4.3 $\pm$ 2.7 cm; range 1-16 cm). All the CT scans were performed using multidetector CT (MDCT) scanners available at Seoul National University Hospital (n = 213) or various MDCT scanners at another hospital (n = 67). The protocol for the liver CT examinations comprised four phases: precontrast, arterial, portal, and delayed phases. After obtaining the precontrast images, 120 mL of 370 mg/mL of iopromide (Ultravist 370; Schering Korea, Seoul, Korea) was injected at a rate of 3.0–4.0 mL/s using a power injector (Envision CT; Medrad, Pittsburgh, PA, USA). Using the bolus tracking method, the arterial phase images were acquired automatically 19 s after an attenuation of 100 Hounsfield units was achieved in the descending aorta. The portal venous and delayed phase images were obtained 33 and 180 s after the arterial phase, respectively, after a contrast agent was administered. The entire dataset was approved by the international review board of the Seoul National University Hospital. The requirement for informed consent was waived.

\subsection{Axial Alignment and Image Preprocessing}
A custom multiphase dataset is generated from axial slices of 3D 512$\times$512 pixel CT scan with variable heights per patient. Similar to the CT image preprocessing pipeline in \cite{christ2016automatic} and \cite{lee2018liver}, the Hounsfield unit range of the raw DICOM images is reduced to [-150, 250] by clipping, which is suitable for detecting hepatic lesions \cite{mayo1999detecting}. After the windowing, the images were normalized to have values between [0, 1] to be used as inputs to the model.

\begin{figure}[t]
\begin{center}
\centerline{\includegraphics[width=0.95\linewidth]{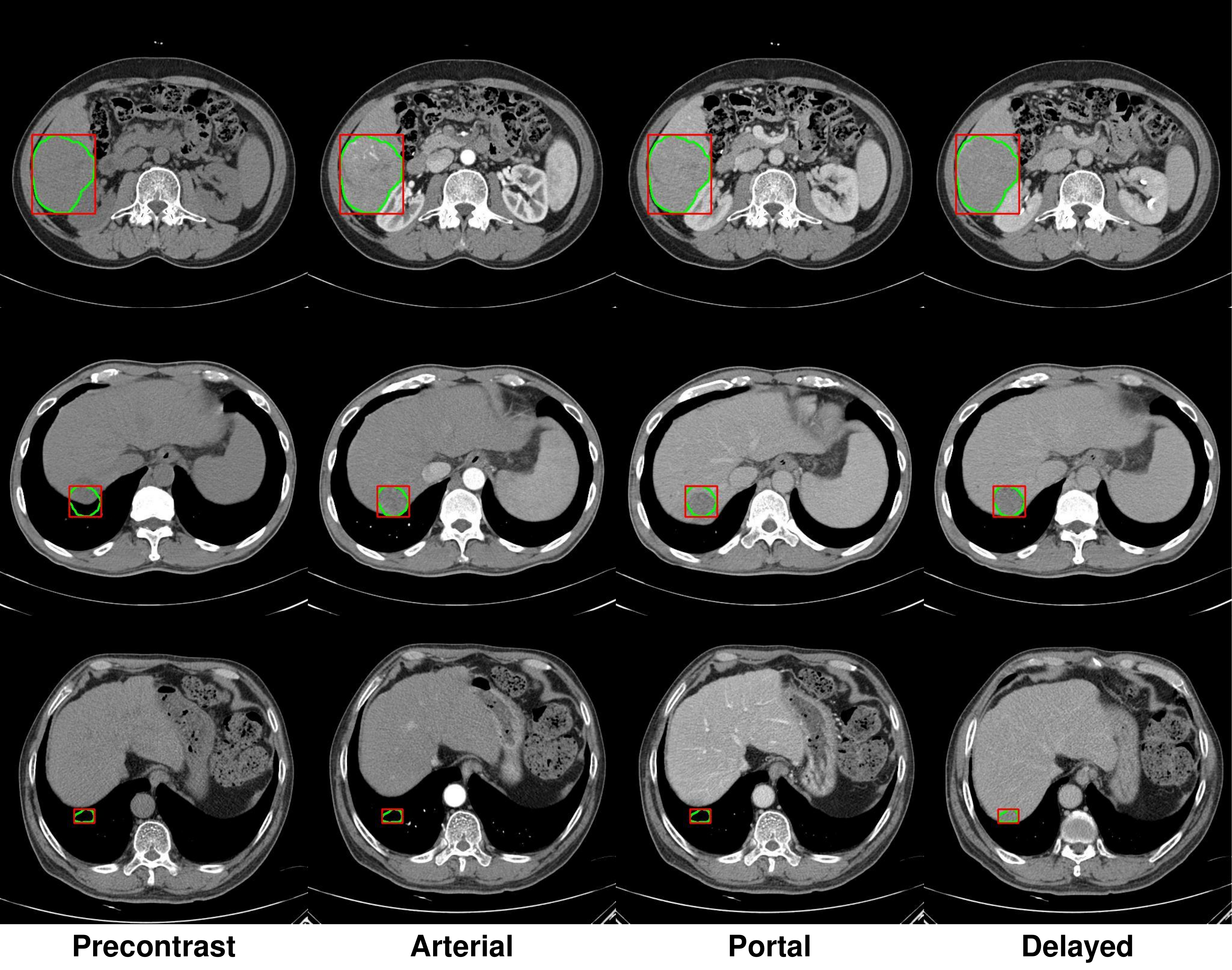}}
\caption{Visualization of custom-designed multiphase focal liver lesion CT dataset. Each row represents a single data point. Green contours correspond to segmentation masks after Gaussian kernel smoothing is applied. Red boxes correspond to bounding box annotations obtained from the segmentation mask for object detection. We annotated the segmentation mask from one of the four phases as the reference. Top: an example of a well-registered case, where the annotation is from the portal phase. Middle and bottom: an example of a poorly registered case, where the annotation is from the delayed phase. Note that the ground truth label is not consistent with the other phases.}
\label{figure_input}
\end{center}
\end{figure}

The axial coordinates of all four phases are manually aligned to ensure that all the phases indicated the same axial location of the 3D CT tensor. Subsequently, three consecutive axial CT slices were stacked (i.e., 2.5D images typically used in deep-learning-based approaches) for each of the four phases as a single data point. This was performed such that the model can leverage the axial features in the original images (e.g., to differentiate between blood vessels and lesions). This resulted in each data point containing a 12-channel image and a total of 4,748 data points. Fig. \ref{figure_input} shows some examples of the preprocessed data.

\begin{figure*}[ht]
\begin{center}
\centerline{\includegraphics[width=0.95\textwidth]{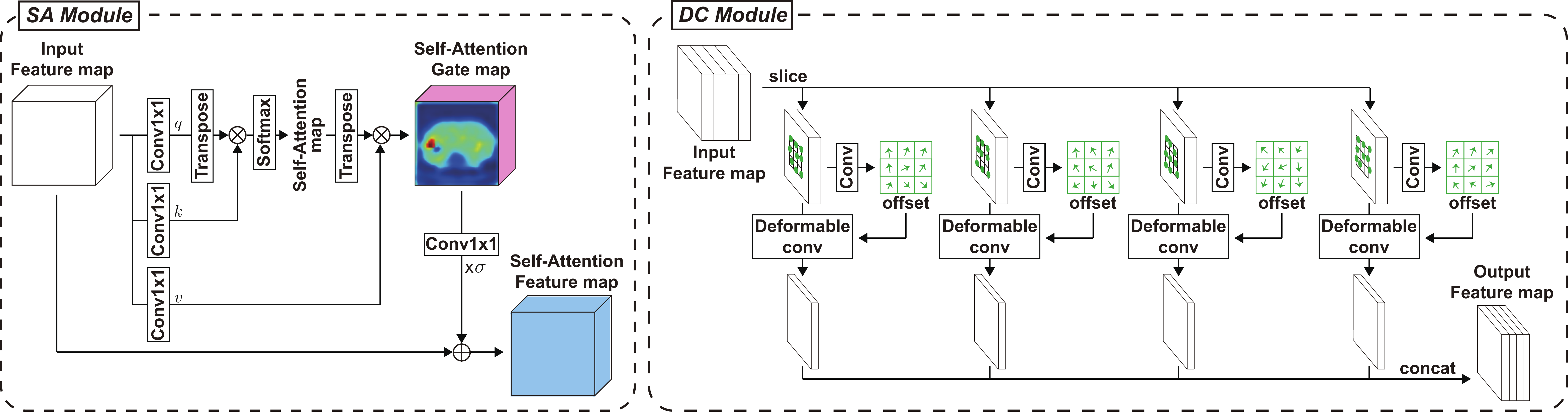}}
\caption{Detailed description of self-attention (SA) module and deformable convolution (DC) module.}
\label{figure_moddule}
\end{center}
\end{figure*}
\subsection{Bounding Box Label Annotation}
A board-certified radiologist manually drew segmentation masks for HCCs on multiphase CT images using a graph cut-based \cite{boykov2006graph} commercially available software (MEDIP; MEDICAL IP, Seoul, Korea). Among the contrast-enhanced phases, the one in which the HCC demonstrated the highest lesion-to-liver contrast was selected for placing the segmentation masks according to the radiologist’s discretion. We opted to construct the following numbers of segmentation masks for the phases: delayed (n = 158), arterial (n = 91), and portal (n = 31).

Besides, a post-annotation filtering with Gaussian blur is applied to the segmentation mask because the graph cut-based segmentation resulted in a fuzzy contour. We observed that a Gaussian kernel measuring $11\times11$  pixels resulted in the highest contour quality. As this study had formulated the task as an object detection problem, the bounding boxes were extracted from the segmentation mask contours using OpenCV \cite{opencv_library} Python package.

As mentioned above, discrepancies were observed in the liver morphology and HCC location among the multiphase images, mainly due to respiratory movement. Consequently, the segmented mask for HCC in a particular phase did not perfectly match the same lesion depicted on the other phases (Fig. \ref{figure_input}). Hence, our dataset posed additional challenges for the model in terms of the phase that the model should learn to select the best phase for the four-phase input to achieve maximal performance.

\section{Method}

This study presents a fully automated end-to-end learning approach for FLL detection from multiphase CT images. The following sections first describe the task-specific functionalities and inductive biases required for achieving maximal model performance on our real-world CT dataset. Subsequently, we present a trainable module that satisfies the desired properties. Finally, we present methods to incorporate the proposed modules into various high-performance single-stage object detection models.

This study was approved by the institutional review board (IRB No. 1704-175-849) of our institution, with a waiver of the requirement for informed consent.

\subsection{Incorporating Parameterized Task-specific Priors}
Standard approaches for deep-learning-based object detection use unimodal images as data distribution, such as a single natural image in the RGB color space. On the contrary, our FLL diagnosis task comprises multiphase images as the input. Hence, the direct application of such models results in significantly degraded performances. In \cite{lee2018liver}, a multiphase variant tailored to liver lesion detection is proposed by administering a grouped convolution~\cite{krizhevsky2012imagenet} to the base CNN and introducing $1\times1$ convolutions on top of the grouped feature map extracted from the base CNN. The detection performance of the aforementioned method is still limited because it does not address the spatial and morphological variances between modalities from the perspective of the model architecture. Specifically, although the $1\times1$ convolution used by the GSSD~\cite{lee2018liver} implicitly conducts the attentive process among the phases, it does not explicitly parameterize the attention mechanism. Moreover, the method is not robust to a registration mismatch between phases because the convolution kernel can apply only a feature transformation on a specified filter space and cannot adjust the spatial mismatch between multiphase features. Our method builds upon the multistream architectural paradigm of GSSD~\cite{lee2018liver} and addresses challenges for further enhancing the performance and achieving robustness to unregistered images.

Hence, this study presents a trainable module that directly incorporates task-specific priors required for real-world data into the neural architecture (Fig.~\ref{figure_moddule}), which denote an adjustment of the spatial mismatch between phases in a deep feature space. This reduces the sensitivity of the detection performance of the model to the quality of the registration. We built the module based upon the strong baseline of the GSSD and introduced several key enhancements to the model and denote it as GSSD++ (Fig. \ref{figure_model}). It is noteworthy that the proposed method is not restricted to the GSSD and is applicable to various single-stage detection models; however, we used the topology of the GSSD architecture for clarity.

The proposed method highlights two main advancements. First, we introduce the dynamic interphase extraction of spatial saliency from multistream features for lesion detection based on the self-attention mechanism~\cite{vaswani2017attention,zhang2018self}. Second, we propose a novel mechanism of attention-guided multiphase spatial alignment based on a deformable convolution kernel~\cite{dai2017deformable, zhu2019deformable}. The following sections provide implementation details of the proposed method and the integration of the modules to high-performance object detection models.

\subsection{Interphase Saliency Extraction with Self-attention}
\label{selfatt}
\begin{figure*}[ht]
\begin{center}
\centerline{\includegraphics[width=0.9\textwidth]{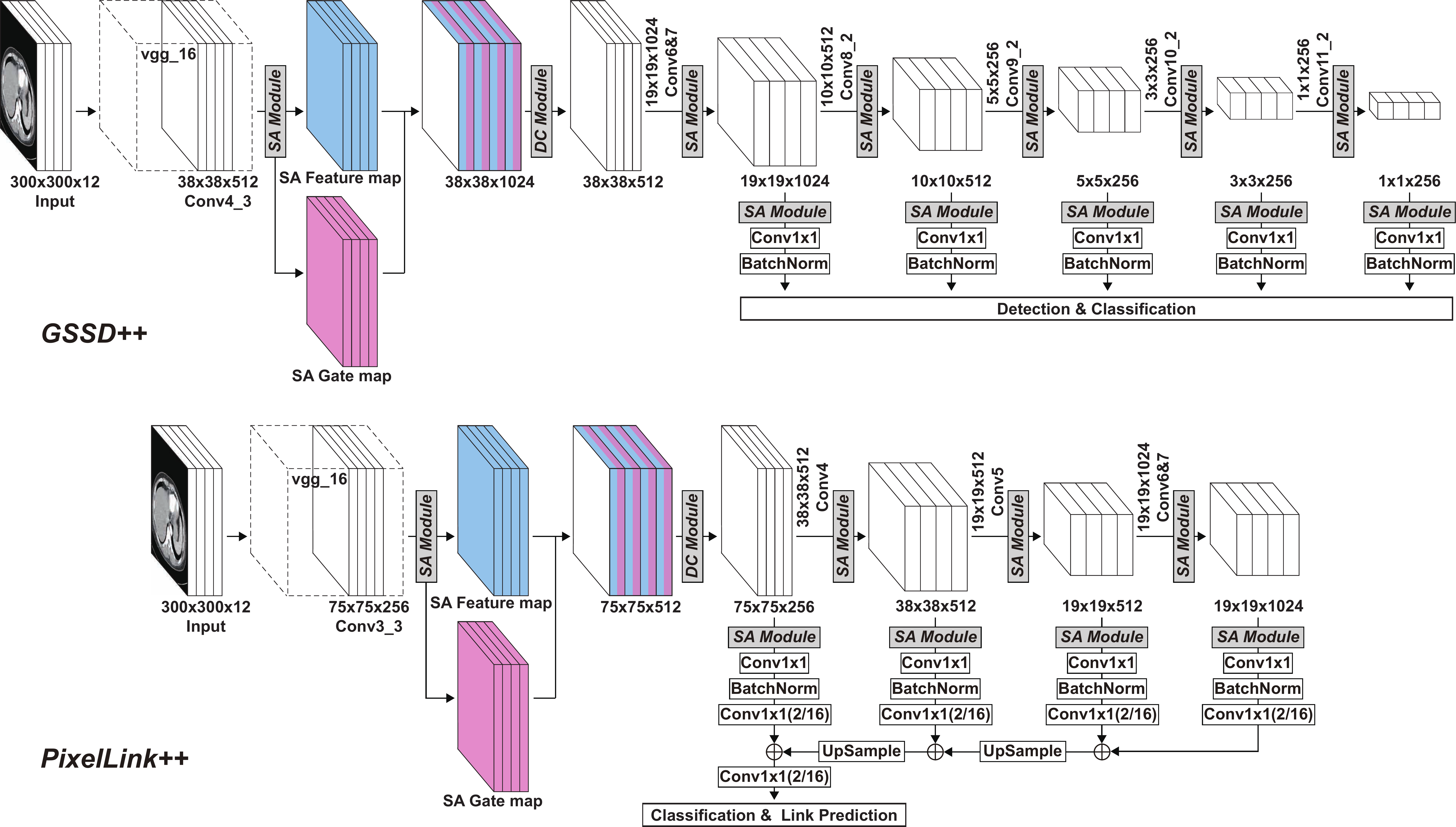}}
\caption{Schematic diagram of GSSD++ (top) and PixelLink++ (bottom). Solid lines drawn along the surface of tensors indicate a feature map constructed by grouped convolutions. SA module refers to self-attention module and DC module refers to deformable convolutions.}
\label{figure_model}
\end{center}
\end{figure*}

This section introduces the self-attention mechanism for extracting interphase spatial saliency. We adopted a modified version of convolutional self-attention layers~\cite{zhang2018self} that operated on the multiphase feature space (Left in Fig. \ref{figure_moddule}). The layer was applied at two locations: one in the grouped feature map from the base CNN, and the other before the multiphase fusion with the $1\times1$ convolution (Fig. \ref{figure_model}). Hence, the grouped feature map of the base CNN can communicate between previously decoupled feature channels inside the self-attention (SA) module. This is further leveraged by global-level spatial dependencies across the feature map from the non-local nature of self-attention. This enables the model to address the salient phase of the multiphase inputs from both the feature extraction and channel fusion stages, thereby enhancing the performance of lesion detection.

Assume that we have a deep image feature map $\bm{x} \in \mathbb{R}^{C\times N}$, where $C$ is the number of feature channels, and $N=H\times W$ is a flattened two-dimensional (2D) image feature. In our model, $C$ was obtained from the concatenated features with $C/4$ channels, each via grouped convolutions. $\bm{x}$ was first transformed into three features, i.e., query $\bm{q}=\bm{W_q}\bm{x}$, key $\bm{k}=\bm{W_k}\bm{x}$, and value $\bm{v}=\bm{W_v}\bm{x}$, where $\bm{W_q} \in \mathbb{R}^{\bar{C}\times C} , \bm{W_k} \in \mathbb{R}^{\bar{C}\times C}$,  and $\bm{W_v} \in \mathbb{R}^{\hat{C}\times C}$ were parameterized by separate $1\times1$ convolutions. $\bar{C}$ and $\hat{C}$ are the number of bottleneck channels during the self-attention, where we used $\bar{C}=C/8$ and $\hat{C}=C/2$. The global-level self-attention map $\bm{\beta} \in \mathbb{R}^{N \times N}$ is obtained as follows:
\begin{equation}
\label{eq1}
\beta_{ji} = \frac{\exp(s_{ij})}{\sum_{i=1}^{N} \exp(s_{ij})}, \quad \text{where} \quad \bm{s} = \bm{q}^{T} \bm{k}.
\end{equation}
The self-attention map $\bm{\beta}$ exhibits global dependencies between features, where $\beta_{ji}$ corresponds to the extent to which a $j^{th}$ feature location attends to an $i^{th}$ location. From the self-attention map, a self-attention gate map 
$\bm{g} \in \mathbb{R}^{\hat{C} \times N}$ is obtained as follows:
\begin{equation}
\label{eq2}
\bm{g} = \bm{v}\bm{\beta}^T.
\end{equation}
Finally, a residual self-attention feature map $\bm{y} \in \mathbb{R}^{C \times N}$ is obtained as follows:
\begin{equation}
\label{eq3}
\bm{y} = \sigma \bm{o} + \bm{x} , \  \text{where} \  \bm{o} = \bm{W_o}\bm{g} \  \text{and} \  \bm{W_o} \in \mathbb{R}^{C \times \hat{C}}.
\end{equation}
$\sigma$ is a trainable parameter that serves as a gating mechanism for residual self-attention. It is initialized to zero such that the architecture can focus on the local dependencies first and then to learn to expand its focus to the non-local features. Consequently, the previously decoupled feature maps can communicate between phases by observing the global context (both channel- and location-wise) via the gated self-attention map, causing the model to adaptively attend to the salient phases and locations. Additionally, the same self-attention module is applied before conducting the $1\times1$ channel fusion convolution to further enhance the self-attention capability.

Furthermore, a spatial bottleneck is applied to the self-attention mechanism based on the target architecture. This enables the model to attend to aggregated pixels instead of individual pixels in the feature space, thereby achieving lower run-time memory consumption and stabilized model dynamics. Specifically, spatial average pooling is applied over the query $\bm{q} \in \mathbb{R}^{\bar{C} \times H \times W}$ to get $\bm{\tilde{q}} \in \mathbb{R}^{\bar{C} \times \frac{H}{D} \times \frac{W}{D}}$ flattened to $\bm{\tilde{q}} \in \mathbb{R}^{\bar{C} \times \frac{N}{D^2}}$, where $D$ is a down-sample factor for the pooling operation. The same pooling was applied to value $\bm{\tilde{v}} \in \mathbb{R}^{\hat{C} \times \frac{N}{D^2}}$ to obtain a gate map with the original spatial resolution $\bm{g} \in \mathbb{R}^{\hat{C} \times N}$ from $\bm{g} = \bm{\tilde{v}}\bm{\tilde{\beta}}^T$, where $\tilde{\beta} \in \mathbb{R}^{N \times \frac{N}{D^2}}$.

\subsection{Attention-guided Multiphase Alignment}
We endowed the system with robustness to registration mismatch by introducing a novel \textit{attention-guided multiphase alignment} module in the feature space. This was achieved by incorporating a modified version of the deformable convolution (dconv)~\cite{dai2017deformable, zhu2019deformable} kernels into the network. We propose applying dconv kernels to the deep grouped feature map extracted by the backbone CNNs, whose alignment operations are guided by the SA module.

\begin{table*}[!t]
\caption{Performance evaluation of various model configurations with average precision on multiple overlap thresholds. Overlap criteria are either intersection over union (IoU) or intersection over bounding box (IoBB). $\dagger$ denotes that segmentation masks are used during training. G.Conv and C.Fusion represent grouped convolutional and $1\times1$ channel fusion layers, respectively. A registration was not conducted on the dataset.}
\label{ap_result}
\centering
  \begin{tabular}{l|ccc|ccc|cc}
    \toprule
    \multirow{2}{*}{Methods} & 
    \multicolumn{6}{c|}{Validation Set(\%)} &
    \multicolumn{2}{c}{Test Set(\%)} \\
    & {IoU30} & {IoU50} & {IoU70} & {IoBB30} & {IoBB50} & {IoBB70} & {IoU50} & {IoBB50} \\
    \midrule
    \midrule
    GSSD \cite{lee2018liver} (Portal Only) & 50.62 & 33.44 & 22.44 & 53.66 & 50.64 & 44.98 & 48.35 & 54.16 \\
    \midrule
    SSD \cite{liu2016ssd} & 55.26 & 43.68 & 36.50 & 57.36 & 55.90 & 52.35 & 58.41 & 63.18 \\
    GSSD \cite{lee2018liver} & 56.58 & 43.72 & 36.41 & 58.23 & 57.18 & 53.07 & 60.06 & 67.33\\
    \midrule
    Improved RetinaNet \cite{zlocha2019improving} $\dagger$ & 51.25 & 42.74 & 22.18 & 51.42 & 47.52 & 41.16 & 49.93 & 53.77\\
    Improved RetinaNet + G.Conv~\&~C.Fusion & 57.72 & 48.63 & 35.21 & 60.08 & 58.51 & 52.48 & 62.47 & 66.30 \\
    Improved RetinaNet + G.Conv~\&~C.Fusion$\dagger$ & 63.46 & 53.11 & 35.67 & 65.21 & 62.85 & 56.64 & 63.46 & 68.13 \\
    \midrule
    PixelLink \cite{deng2018pixellink} & 61.25 & 45.81 & 15.14 & 65.23 & 52.86 & 24.59 & 48.44 & 56.05 \\
    PixelLink + G.Conv~\&~C.Fusion & 69.20 & 50.85 & 19.90 & 71.91 & 56.08 & 29.64 & 57.89 & 64.44 \\
    MSCR \cite{liang2019multi} & 63.72 & 51.33 & 28.14 & 67.10 & 57.50 & 41.52 & 55.57 & 60.02 \\
    \midrule
    GSSD++ (ours) & 62.41 & 52.90 & \textbf{40.08} & 67.89 & 65.20 & \textbf{59.99} & \textbf{67.87} & \textbf{77.22} \\
    PixelLink++ (ours) & \textbf{74.12} & \textbf{57.77} & 24.01 & \textbf{77.66} & \textbf{65.30} & 33.99 & 64.19 & 73.52 \\
    \bottomrule
  \end{tabular}
\end{table*}

The dconv kernels have a trainable and dynamic receptive field based on inputs obtained by predicting the offsets to the original sampling grid by a separate convolutional layer. Subsequently, the deformable convolution filter operates on the augmented grid predicted by the offsets. Concretely, define the regular sampling grid $\mathcal{R}$ of a $3\times3$ convolution filter with dilation 1 as $\mathcal{R} = \{(-1, -1), (-1, 0), ..., (0, 1), (1, 1)\}$. The regular convolution filter on this fixed grid $\mathcal{R}$ operates as follows:
\begin{equation}
\label{eq4}
\bm{y}(\bm{p_0}) = \sum_{\bm{p_n \in \mathcal{R}}} \bm{w}(\bm{p}_n) \cdot \bm{x}(\bm{p}_0 + \bm{p}_n),
\end{equation}
where $\bm{p}_0$ is the output location from the convolution filter $\bm{w}$ applied to input $\bm{x}$.

The offset predictor computes $\{\Delta\bm{p}_{n|n=1,...,N}\}$, where $N=|\mathcal{R}|$, through the separate convolution layer. Subsequently, the deformable convolution filter operates on the augmented grid as follows:

\begin{equation}
\label{eq5}
\bm{y}(\bm{p_0}) = \sum_{\bm{p_n \in \mathcal{R}}} \bm{w}(\bm{p}_n) \cdot \bm{x}(\bm{p}_0 + \bm{p}_n + \Delta \bm{p}_n).
\end{equation}

The conventional application of the dconv layer predicts the same offsets across all channels. In contrast, because our desired functionality was the interphase feature-space alignment, the deformable convolution (DC) module is tasked to predict the four separate offsets corresponding to each grouped feature map. Subsequently, we applied dconv kernels with multiple offsets targeted differently for each grouped feature channel (Right in Fig. \ref{figure_moddule}).

The DC module is inserted after the self-attended feature map of the first source layers (e.g., conv4\_3 of the VGG16~\cite{simonyan2014very} architecture in the GSSD model). The main objective is to have the model adaptively align the mismatch of offsets and morphological variances between phases from a fine-grained grouped feature map earlier in the base CNN forward pass. The DC module should receive a grouped feature map with a deep representation that is sufficiently rich to fully leverage the dconv filters and avoid overfitting. We observed that inserting the AC and DC modules into conv4\_3 yielded the best results, i.e., better loss and detection metrics, while simultaneously minimizing overfitting.

The dconv kernels with multiphase offset predictors require relevant visual cues to possess the functionality of the intermodal alignment. This is achieved by guiding the learning process of the dconv kernels by leveraging the modified self-attention module described in the preceding section. Contrary to~\cite{zhang2018self}, where the method forward-propagates only the self-attention feature map, we propose to further leverage the self-attention gate map as an additional visual cue for the attention-guided multiphase alignment. Specifically, the gated self-attention map is concatenated with the feature map and processed by the DC module (Fig. \ref{figure_model}). By receiving both the features and self-attention gate map as the inputs, the dconv filters can adaptively predict the geometric offsets for the multiphase features, with the global-level self-attention map as an additional context. The self-attention gate map can additionally receive gradients from the dconv layers directly, providing a shortcut for cooperative learning between two modules. We observed that this novel combination significantly improved the detection performance on the misregistered multiphase dataset.

\section{Experimental results}
The proposed method can be easily applied to various single-stage detection networks. To demonstrate its applicability, this study applied the proposed modules to the GSSD \cite{lee2018liver} and PixelLink~\cite{deng2018pixellink}, which are referred to as GSSD++ and PixelLink++, respectively; subsequently, we performed a comparative and ablative analysis using recent state-of-the-art FLL detection methods \cite{lee2018liver,liang2019multi,zlocha2019improving}.

\begin{figure*}[t]
\centering
\centerline{\includegraphics[width=0.8\textwidth]{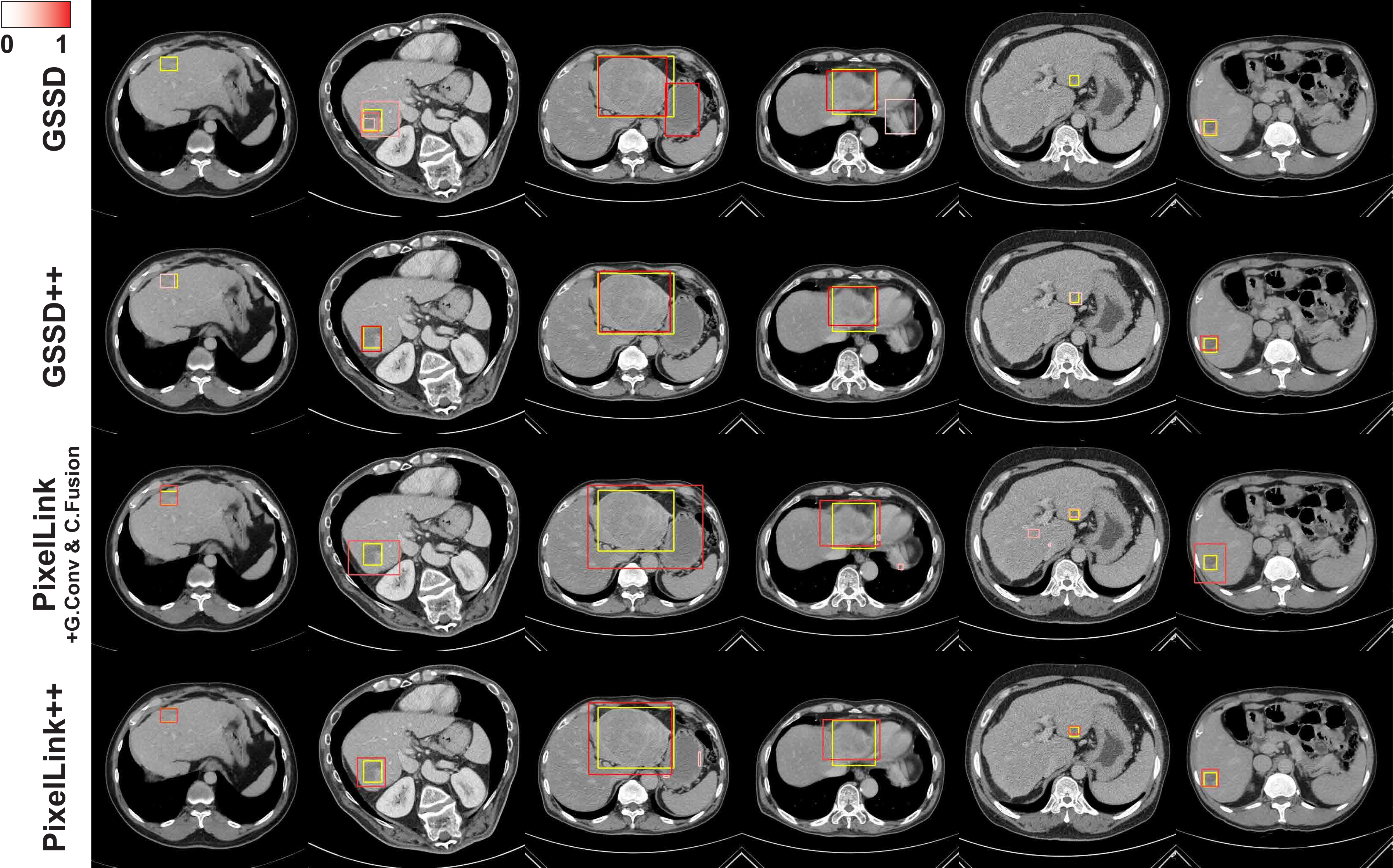}}
\caption{Qualitative evaluation of liver lesion detection performance between models. Yellow rectangle corresponds to the ground truth. Box predictions from the model are colored ranging between white to red which corresponds to the classification confidence ranging from 0 to 1.}
\label{figure_compare}
\end{figure*}

\subsection{Implementation Details}
We used VGG16~\cite{simonyan2014very} as their backbone CNN architecture to accurately assess the performance of each model by decoupling the effectiveness of the proposed modules from incorporating more powerful base CNN architectures~\cite{he2016deep}. The downsample factor $D$ for the pooling operation inside the self-attention mechanism in Section \ref{selfatt} is applied as $D=1$ for GSSD++ and $D=2$ for PixelLink++ from the hyper-parameter sweep over $D=\{1, 2, 4, 8\}$. For evaluation, the custom-curated 280-patient dataset is randomly split into 216, 54, and 10 patients and used for training, validation, and testing, respectively.

Similar to \cite{liu2016ssd} and \cite{lee2018liver}, we applied most of the on-the-fly data augmentation methods from the vanilla SSD training (including random crop, mirror, expansion, and shrink) together with brightness and contrast randomization. The training loss terms defined in each model are used without modification. A positive:negative ratio of 1:3 is used for the online hard negative mining \cite{liu2016ssd} training method, except for Improved RetinaNet~\cite{zlocha2019improving}, which used the focal loss~\cite{lin2017focal}.

\subsection{Evaluation Metrics}
We quantitatively measured the performance of each model based on the average precision (AP) over either the intersection over union (IoU) or intersection over bounding box (IoBB) \cite{wang2017chestx}, which is a standard benchmark for deep-learning-based object detection models on natural images. The low-confidence predictions below 0.2 are filtered out to focus on the lesion detection performance of high-confidence box predictions and compare the performance of various models fairly.

\subsection{Baseline Results}
We measured the effectiveness of using multiphase CT images by training the baseline model using only the portal phase images (Table \ref{ap_result}). The model performed significantly worse when trained with only portal phase images than with the multiphase baseline. This verifies that the GSSD leverages multiphase data effectively, consequently verifying the requirement for robustness to the multiphase registration mismatch of the model. Because the original architecture from the improved RetinaNet~\cite{zlocha2019improving} and PixelLink~\cite{deng2018pixellink} do not consider multiphase data, we adopted a grouped convolutional layer and a $1\times1$ convolutional channel fusion layer introduced in \cite{lee2018liver} to those models and confirmed that the method enabled the model to leverage multiphase data more effectively.

RetinaNet~\cite{lin2017focal} and its optimized variant for liver lesion detection \cite{zlocha2019improving} are based on Feature Pyramid Networks (FPNs)~\cite{lin2017feature}. Although the model is often considered as a state-of-the-art model, we did not observe any significant performance gain compared with other baselines when training was performed using bounding box annotations. The model demonstrated a clear performance improvement only when a dense segmentation mask label was used additionally (instead of the RECIST-based mask label used by \cite{zlocha2019improving}, which is not accessible in our data). We refer to the model trained with stronger supervision of the segmentation mask as a strong baseline in our experimental result.

PixelLink \cite{deng2018pixellink} cast the bounding box regression problem into instance segmentation and used a balanced cross-entropy loss for a small area, enabling PixelLink-based models to excel at detecting smaller-sized lesions from the validation set. However, owing to the noisy post-filtering approach for the bounding box extraction from the predicted segmentation map, the performance degraded for metrics with a higher overlap threshold. SSD \cite{liu2016ssd} and RetinaNet \cite{lin2017focal}, which used a more conventional approach involving the use of a predefined set of densely distributed anchor boxes \cite{ren2015faster}, achieved better performances on metrics with higher thresholds.

\subsection{Improvements from the proposed method}
Table \ref{ap_result} summarizes the detection performance of the proposed framework, GSSD++ and PixelLink++. As shown, the proposed models significantly outperformed the baseline models for all the metrics with various overlap thresholds. GSSD++ achieved an AP of 77.22\% on IoBB50, i.e., 9.89\%p higher than the GSSD. PixelLink++ achieved an AP of 73.52\% on IoBB50, i.e., 9.08\%p higher than PixelLink with grouped convolution and a channel fusion layer.
Fig.~\ref{figure_compare} shows a qualitative evaluation of the liver lesion detection performances of GSSD++ and PixelLink++ compared with the baselines. Our models detected lesions with higher confidence (marked in red) and indicated fewer false positives compared with the baseline models, which localized the wrong location as the lesion.
For $2^{nd}, 3^{rd}$, and $6^{th}$ columns in Fig.~\ref{figure_compare}, the baseline models detected an area as a lesion larger than the actual lesion. As shown in Fig.~\ref{figure_input}, a spatial mismatch occured between the phases, which made it difficult for models without an internal mismatch ability to find accurate lesion regions.

\subsection{Robustness to registration mismatch}\label{sec:robust}
To verify the robustness of the proposed method to a registration mismatch, we trained all models using the registered dataset and measured the performance degradation owing to such a mismatch.
The data is registered using three different methods: rigid, affine, and non-rigid, by using the SimpleElastix~\cite{marstal2016simpleelastix} open-source package with the default parameters and considering the portal phase as a fixed image\footnote{It took 30 seconds for the non-rigid registration of a single CT slice among the four phases on two Intel Xeon E5-2690 v4 CPUs with 2.6 GHz and 56 threads.}. Each 2D slice image and lesion segmentation mask was registered on the portal phase image, and the bounding box was then extracted from the registered segmentation mask contours. Table~\ref{tab:mismatch} shows the level of registration mismatch of each dataset, which was measured with an overlap of the liver obtained from a liver segmentation network\footnote{https://github.com/andreped/livermask}. Motivated by \cite{balakrishnan2018unsupervised}, we calculated the Dice score~\cite{dice1945measures} between the liver from two different phases. For reliability, the Dice scores were averaged except for those cases in which the values were zero.

We defined the sensitivity of each model configuration to a registration mismatch as the ratio of performance degradation in the unregistered dataset when compared with the performance on each registered dataset:
\begin{equation}
\label{eq:robustness}
\text{sensitivity} = 1 - \frac{\text{perf.}(f_\text{unregistered},~\mathcal{D}_\text{unregistered})}{\text{perf.}(f_\text{R},~\mathcal{D}_\text{R})},
\end{equation}
where $\mathcal{D}_\text{unregistered}$ is the unregistered dataset and $\mathcal{D}_R$ is the dataset registered with method $R$. In addition, $\text{perf.}(f_R,~\mathcal{D}_R)$ is the performance of model $f$ on dataset $\mathcal{D}_R$, which is trained with dataset $\mathcal{D}_R$.

\begin{table}[t]
  \centering
  \caption{Degree of mismatch based on liver.}
  \vspace{-0.5em}
\centering
  \begin{tabular}{l|c}
    \toprule
    Methods & Dice Score\\ 
    \midrule
    Unregistered	& 0.9001 \\
    Rigid	& 0.9119 \\
    Affine	& 0.9213 \\
    Non-rigid	& 0.9622 \\
    \bottomrule
  \end{tabular}
  \label{tab:mismatch}%
\end{table}%

\begin{figure}[t]
\centering
\centerline{\includegraphics[width=0.95\columnwidth]{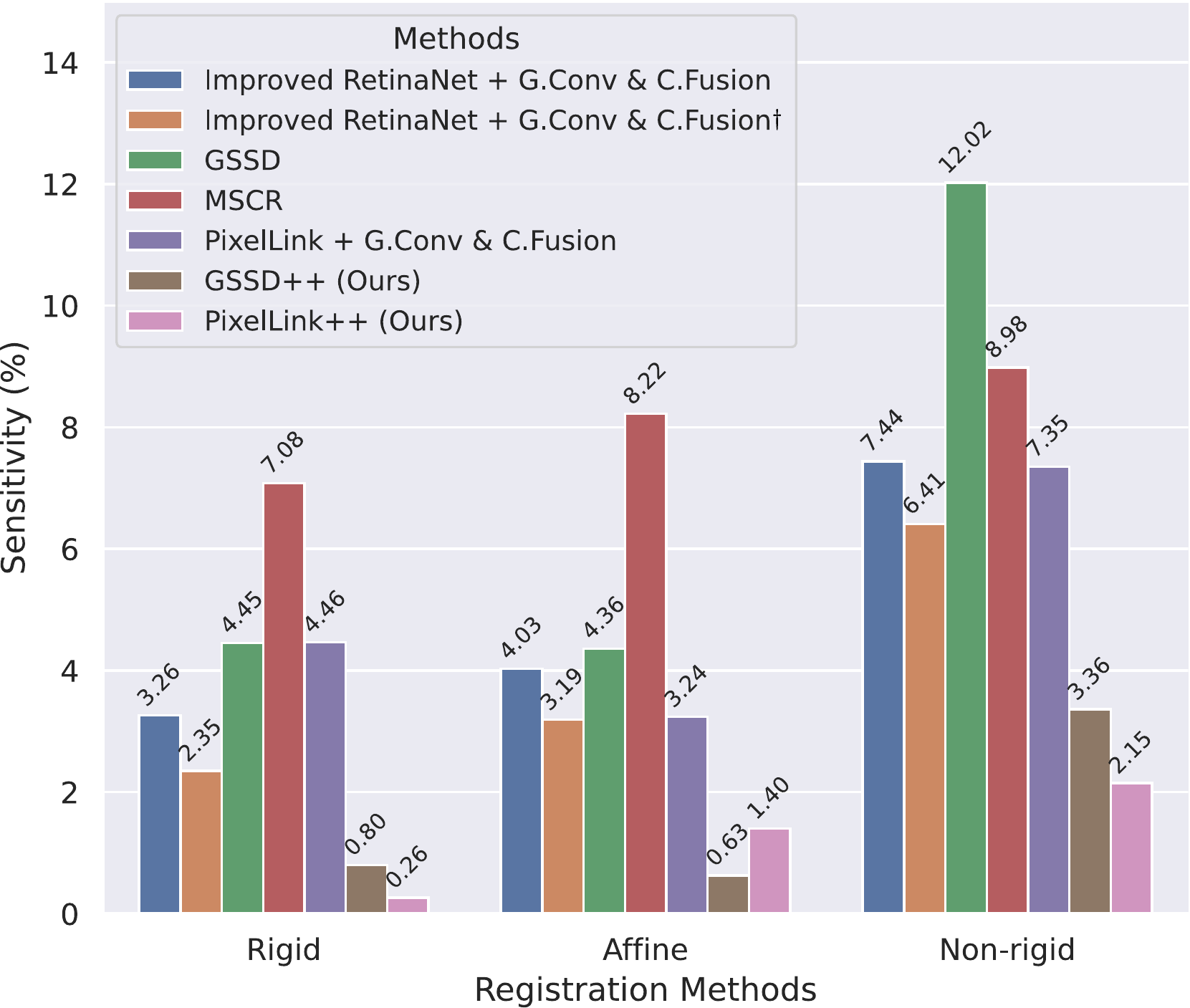}}
\caption{Assessment of robustness to registration mismatch. Each model was trained using the registered version of a multiphase dataset by applying various methods from SimpleElastix. Here, $\dagger$ denotes that segmentation masks were used during training. A lower sensitivity indicates higher robustness to a registration mismatch.}
\label{figure_sensitivity}
\end{figure}

Fig.~\ref{figure_sensitivity} shows the average sensitivity of each model configuration for various levels of mismatch based on all metrics reported in Table~\ref{ap_result} (AP with IoU30, IoU50, IoU70, IoBB30, IoBB50, and IoBB70 on the validation set, and IoU50 and IoBB50 on the test set).
Because the registered data alleviated the modeling difficulties for the task, all models exhibited a performance degradation (positive sensitivity). Because the dataset registered with a non-rigid method was the best aligned dataset (Table~\ref{tab:mismatch}), the sensitivities of all models on $\mathcal{D}_\text{non-rigid}$ were larger than those on $\mathcal{D}_\text{affine}$ and $\mathcal{D}_\text{rigid}$.
A model can be regarded as more \textit{robust to} a registration misalignment if it has \textit{low} sensitivity owing to \textit{less} degradation in its performance.
We discovered that the sensitivities of GSSD++ and PixelLink++ were significantly lower than those of the baseline models. The results verified that the proposed method is more generalizable than previous approaches owing to its robustness to a registration misalignment between phases. The detection performance of each model is reported in the supplementary material.

\begin{figure}[t]
\begin{center}
\centerline{\includegraphics[width=0.95\columnwidth]{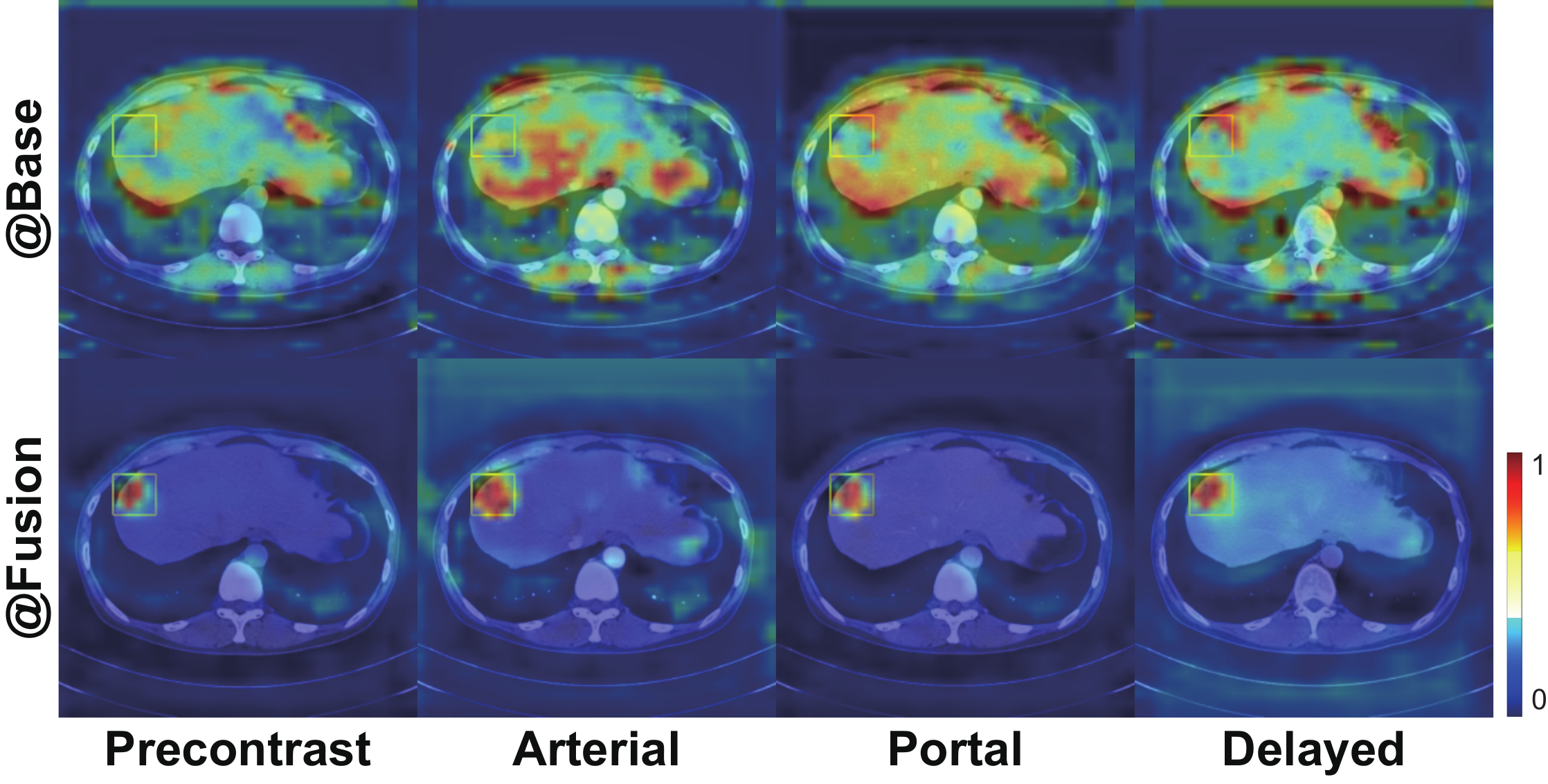}}
\caption{Visualization of self-attention gate maps. Yellow rectangle indicates a ground truth bounding box.}
\label{figure_attn}
\end{center}
\end{figure}
\begin{figure}[t]
\begin{center}
\centerline{\includegraphics[width=0.95\columnwidth]{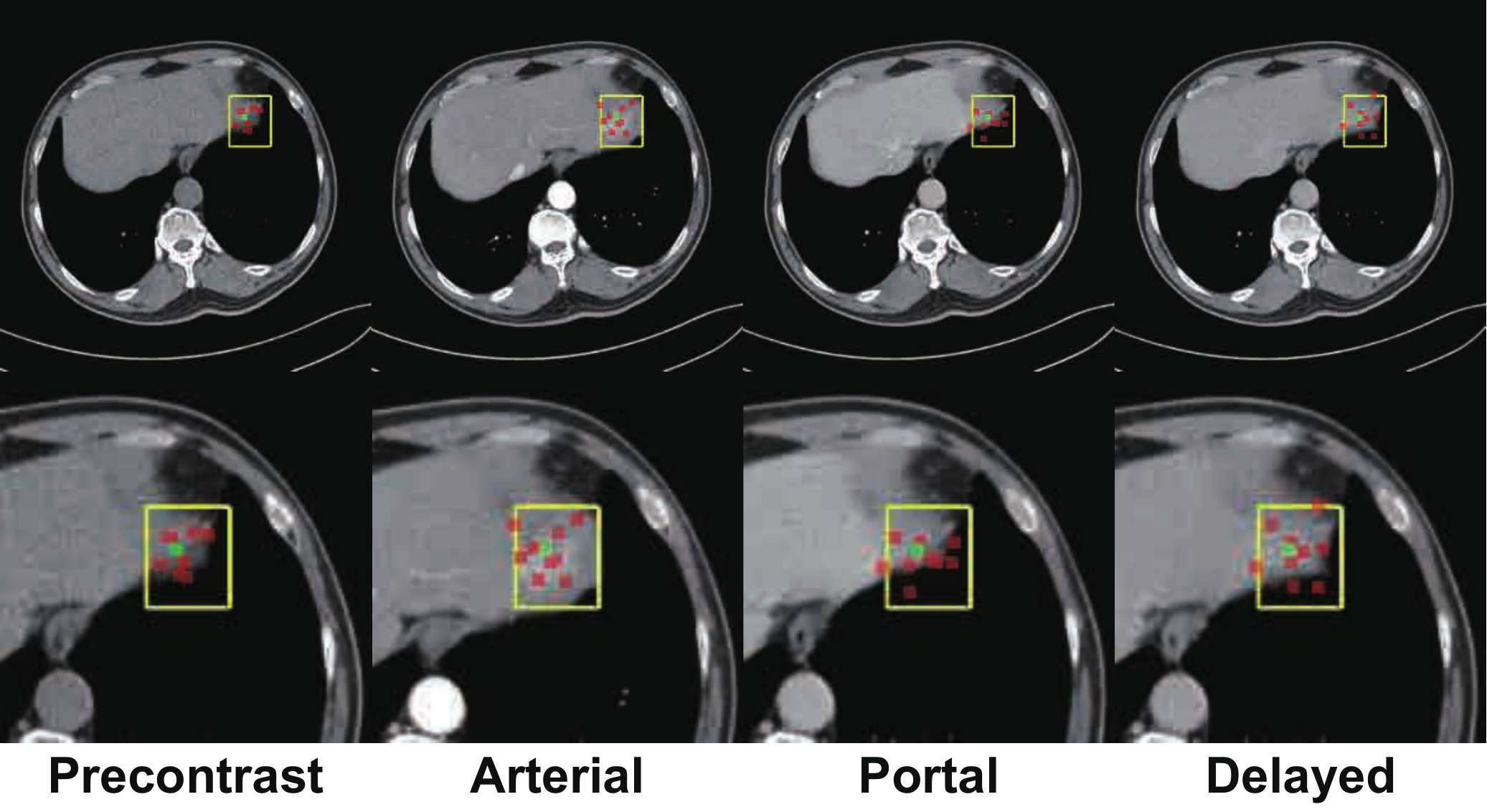}}
\caption{
Visualization of deformable convolution (dconv) filters in a DC module. Yellow rectangle indicates a ground truth bounding box. Green dot refers to a convolution output point in a dconv sampling grid. Red dots correspond to the augmented receptive field for the dconv layer (nine dots per phase for a $3\times3$ kernel). Second row in each example provides a zoomed-in view.}
\label{figure_dcn}
\end{center}
\end{figure}

\subsection{Qualitative Analysis of Model Dynamics}
We qualitatively examined the model behavior of GSSD++ by visualizing a self-attention map and the augmented receptive fields of the dconv kernels. Fig.~\ref{figure_attn} shows the self-attention gate map overlaid on the inputs. As shown, the SA module functioned according to its location based on the different inductive biases encouraged by the model. In the base architecture, the gate map augmented by the self-attention weight was primarily focused on the shape of the organ, thereby providing enhanced visual cues for multiphase alignment and extracting saliency from the organ. In the fusion layer, self-attention successfully localized the salient aspects of the lesion by focusing on the characteristic textures present in the HCCs. The self-attention gate map is computed in a channel-wise manner from the global-level self-attention map ($\bm{\beta}$ in Equation~\ref{eq1}). This can be augmented based on the context present on each feature channel as well as guide the learning of an intermodal alignment and saliency extraction of lesions in an adaptive manner.

Fig.~\ref{figure_dcn} shows the dynamically augmented, phase-wise sampling grid of the dconv kernel. Here, the model adjusted the offsets for the precontrast and portal phase kernels in the left and right examples for the heavily deformed area of the liver, respectively, to align the multiphase deep features across phases. By adaptively correcting the registration mismatches inherent to the dataset guided by self-attention, the model can extract aligned deep feature maps more accurately and thereby achieve a robust object detection.

\subsection{Ablation Analysis}
We performed an ablation analysis of GSSD++ and PixelLink++ to clearly demonstrate the superior performance yielded by the proposed modules. Compared with the final model configuration, modified variants obtained by removing certain modules from the final model indicated degraded performances with various degrees (Table \ref{ablation}).

The dynamic phase-wise offset predictor, which allowed the features of each phase to be aligned, aided increase the detection performance in comparison with that afforded by a global offset predictor. Completely removing the DC module degraded the performance, which deteriorated even further because the model became more inclined to operate on a regular convolution sampling grid rather than a dynamic one. This shows the effectiveness of the alignment in the feature space through the DC module. To create the model variants without an interphase saliency extraction, we applied grouped convolutions~\cite{krizhevsky2012imagenet} in the SA module (``w/o Interphase Attention'' in Table \ref{ablation}). Removing the interphase saliency extraction resulted in no information exchange between phases, which degraded the performance. This shows the effectiveness of saliency extraction through interphase communication.
When the model lost its self-attention capability by the removal of corresponding modules (on both the base CNN and channel fusion locations), the performance was negatively affected. Even when a dynamic sampling grid was present, the model had to detect lesions from the equally weighted feature map without attending to the salient aspects, thereby resulting in a significant performance loss.
The ablation result further highlights the importance of the novel method for attention-guided intermodal alignment to achieve the best lesion detection performance.

\begin{table}[t]
\caption{Ablation analysis of GSSD++ and PixelLink++. Each row corresponds to a modified model by disabling certain modules from the entire model. A registration was not conducted on the dataset.}
\label{ablation}
\centering
  \begin{tabular}{l|cc|cc}
    \toprule
    \multirow{2}{*}{Methods} &
    \multicolumn{2}{c|}{Validation Set(\%)} &
    \multicolumn{2}{c}{Test Set(\%)}\\
    & {IoU50} & {IoBB50} & {IoU50} & {IoBB50} \\
    \midrule
    \midrule
    GSSD++ & \textbf{52.90} & \textbf{65.20} & \textbf{67.87} & \textbf{77.22} \\
    \midrule
    w/o Phase-wise Offsets &  50.92	& 64.44 & 61.50 & 74.08 \\
    w/o DC Module & 51.41	& 62.70 & 61.60 & 71.89 \\
    w/o SA Modules@Base & 52.62	& 64.93 & 62.59 & 74.40 \\
    w/o SA Modules@Fusion & 52.72	& 62.71 & 64.03 & 76.49 \\
    w/o Interphase Attention & 52.49 & 63.14 & 63.82 & 73.56\\
    \midrule
    \midrule
    PixelLink++ & \textbf{57.77} & \textbf{65.30} & \textbf{64.19} & \textbf{73.52} \\
    \midrule
    w/o Phase-wise Offsets &  56.34	& 63.25 & 61.81 & 69.43 \\
    w/o DC Module & 52.24	& 60.12 & 54.27 & 62.14 \\
    w/o SA Modules@Base & 53.70	& 60.22 & 58.39 & 67.43 \\
    w/o SA Modules@Fusion & 54.70	& 60.40 & 59.06 & 67.23 \\
    w/o Interphase Attention & 52.85	& 59.17 & 59.40 & 67.46 \\
    \bottomrule
  \end{tabular}
\end{table}
\section{Conclusion}

This study presented a fully automated, end-to-end liver lesion detection model that leveraged high-performance deep-learning-based single-stage object detection models. The proposed method was achieved by directly incorporating task-specific functionalities and inductive biases as parameterized modules into the model architecture, utilizing interphase self-attention and phase-wise deformable convolution kernels.
We demonstrated a significantly improved detection performance by applying the proposed methods to two fundamentally different approaches for object detection, namely the GSSD \cite{lee2018liver} and PixelLink \cite{deng2018pixellink}; this implies that the proposed method is transferable to various single-stage detection models.
Extensive experiments on both unregistered and registered real-world CT data revealed that the model outperformed previous state-of-the-art lesion detection models through inductive biases targeted at challenges inherent to real-world multiphase data.

The method is applicable to real-world multiphase CT data because the model does not require a perfect registration, which complicates the pipeline. The reduced sensitiveness and reliance on external registration methods would render the deployment of the model into real-world environments significantly more manageable. We believe that simplifying the process towards an end-to-end and scalable approach would enhance the clinical adoption of the deep-learning-based computer-aided detection system.


%


\ifCLASSOPTIONcaptionsoff
  \newpage
\fi



%



\bibliographystyle{IEEEtran}
\bibliography{8.Refs}

%
\clearpage
\begin{IEEEbiography}[{\includegraphics[width=1in,height=1.25in,clip,keepaspectratio]{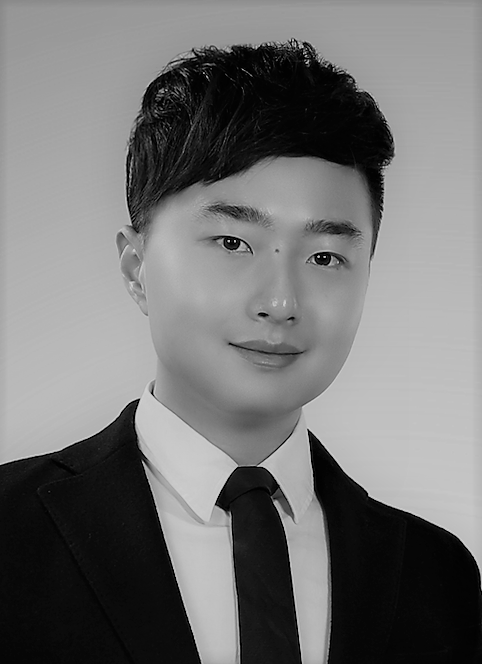}}]{Sang-gil Lee} received the dual B.S. degree in electrical and computer engineering, and applied biology and chemistry from Seoul National University, Seoul, South Korea, in 2016, where he is currently pursuing the integrated M.S./Ph.D. degree in electrical and computer engineering. His research interests include artificial intelligence, deep learning, and medical image computing.
\end{IEEEbiography}

\vskip -2\baselineskip plus -1fil

\begin{IEEEbiography}[{\includegraphics[width=1in,height=1.25in,clip,keepaspectratio]{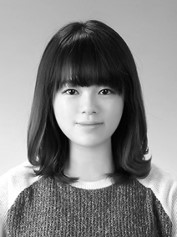}}]{Eunji Kim}
received the B.S. degree in electrical and computer engineering from Seoul National University, Seoul, South Korea, in 2018, where she is currently pursuing the integrated M.S./Ph.D. degree in electrical  and  computer  engineering. Her research interests include artificial intelligence, deep learning, and computer vision.
\end{IEEEbiography}

\vskip -2\baselineskip plus -1fil

\begin{IEEEbiography}[{\includegraphics[width=1in,height=1.25in,clip,keepaspectratio]{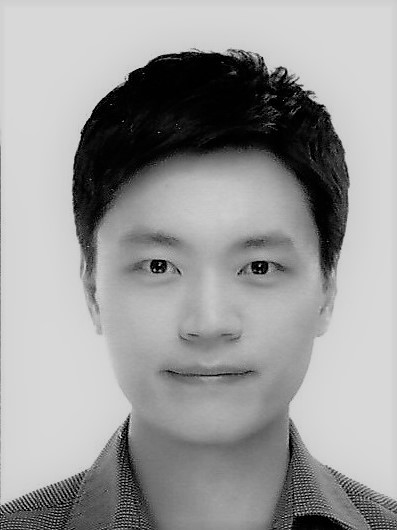}}]{Jae~Seok~Bae}
received the B.S. degree in biological sciences from Korea Advanced Institute of Science and Technology, South Korea, in 2008. He received the M.D. degree from the College of Medicine, Seoul National University, South Korea, in 2012, and M.S. and Ph.D. degrees from Seoul National University, in 2017 and 2020, respectively.
He is currently a Clinical Assistant Professor with the Department of Radiology, Seoul National University Hospital.
\end{IEEEbiography}

\vskip -2\baselineskip plus -1fil
\newpage
\begin{IEEEbiography}[{\includegraphics[width=1in,height=1.25in,clip,keepaspectratio]{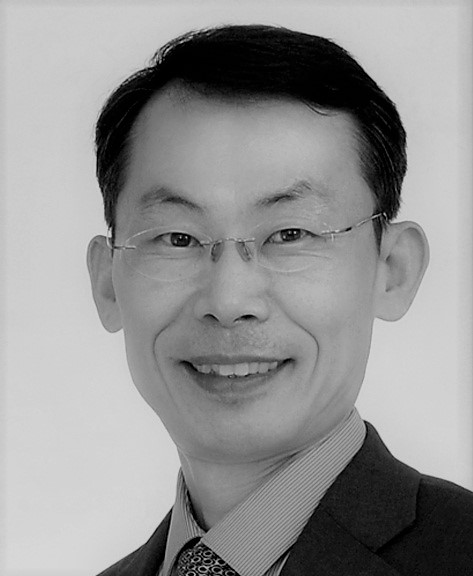}}]{Jung~Hoon~Kim}
received the M.D. degree from the College of Medicine, M.D. degree from the College of Medicine, Hanyang University, South Korea, in 1994, and M.S./Ph.D. degree from Catholic University, South Korea, in 2005.
He was an Assistant Professor with the Department of Radiology, Soonchunhyang University Hospital, South Korea, from 2005 to 2008, and a Visiting Professor and Visiting Assistant Professor with the Department of Radiology and Body Imaging, University of Iowa Hospitals \& Clinics, Iowa City, from 2008 to 2009.
He is currently a Professor with the Department of Radiology and Institute of Radiation Medicine, Seoul National University College of Medicine.

\end{IEEEbiography}

\vskip -2\baselineskip plus -1fil

\begin{IEEEbiography}[{\includegraphics[width=1in,height=1.25in,clip,keepaspectratio]{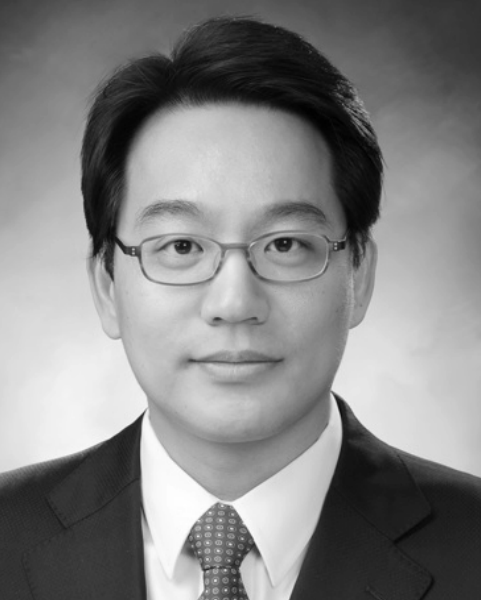}}]{Sungroh Yoon}
(S’99–M’06–SM’11)
received the B.S. degree in electrical engineering from Seoul National University, South Korea, in 1996, and the M.S. and Ph.D. degrees in electrical engineering from Stanford University, CA, USA, in 2002 and 2006, respectively, where he was a Visiting Scholar with the Department of Neurology and Neurological Sciences, from 2016 to 2017.
He held research positions at Stanford University and Synopsys, Inc., Mountain View. From 2006 to 2007, he was with Intel Corporation, Santa Clara.
He was an Assistant Professor with the School of Electrical Engineering, Korea University, from 2007 to 2012. He is currently a Professor with the Department of Electrical and Computer Engineering, Seoul National University, South Korea.
His current research interests include machine learning and artificial intelligence.
He was a recipient of the SNU Education Award, in 2018, the IBM Faculty Award, in 2018, the Korean Government Researcher of the Month Award, 2018, the BRIC Best Research of the Year, in 2018, the IMIA Best Paper Award, in 2017, the Microsoft Collaborative Research Grant, in 2017 and 2020, the SBS Foundation Award, in 2016, the IEEE Young IT Engineer Award, in 2013, and many other prestigious awards.
Since February 2020, Prof. Yoon has been serving as the chairperson (minister-level position) of the Presidential Committee on the Fourth Industrial Revolution established by the Korean Government. 
\end{IEEEbiography}



\clearpage
\setcounter{section}{0}
\renewcommand\thesection{\Alph{section}}
\setcounter{table}{0}
\renewcommand{\thetable}{S\arabic{table}}
\setcounter{figure}{0}
\renewcommand{\thefigure}{S\arabic{figure}}
\noindent {\huge Appendix}
\section{Details in 3D CT Data Collection}

All the CT scans were performed using multidetector CT scanners available at Seoul National University Hospital (n= 213): 320-MDCT (Aquilion ONE, Canon Medical Systems, Tokyo, Japan) (n = 9), 64-MDCT (Brilliance 64, Ingenuity, or IQon, Philips Healthcare, Cleveland, OH, USA; Somatom Definition, Siemens Healthineers, Forchheim, Germany; or Discovery CT750 HD, GE Healthcare, Milwaukee, WI, USA) (n = 131), 16-MDCT (Sensation 16, Siemens Healthineers, Forchheim, Germany; or Mx 8000 IDT 16, Philips Healthcare, Cleveland, OH, USA) (n = 68), 8-MDCT (LightSpeed Ultra, GE Healthcare, Milwaukee, WI, USA) (n = 5), or various MDCT scanners at another hospital (n = 67). Because our data were retrospectively collected from a tertiary referral hospital, the images were obtained from the CT machines of various manufacturers. However, all CT examinations were performed using the routine protocol for liver CT, including those performed at other hospitals.

\section{Details in Training}
We trained all models with a batch size of 32 until convergence was attained. We employed a stochastic gradient descent optimizer with a momentum of 0.9 and weight decay of $5\times10^{-4}$ to train the SSD- and PixelLink-based models. The models were trained for 60,000 training iterations, with a learning rate of $10^{-3}$. We reduced the learning rate by 0.1 times after 30,000 and 50,000 iterations. Furthermore, we reduced the learning rate for the dconv kernel by 0.1 to ensure numerical stability throughout the training process \cite{zhu2019deformable}. 
The improved RetinaNet~\cite{zlocha2019improving} and MSCR~\cite{liang2019multi} were trained based on the protocol of the corresponding open-source implementations\footnote{We used other training configurations as well, including ours, on these models; however, worse performances were obtained.}.

We implemented the model using the PyTorch library and trained the models on an NVIDIA V100 GPU with 32 GB of VRAM.

\section{Data Split}
To perform a fair and unbiased model evaluation without data leakage, we split our custom dataset prior to the experiment without any manual data examination. Consequently, the values of the test set metrics were higher than those of the validation set metrics across all the experiments (IoU50 and IoBB50 in Tables \ref{ap_result}, \ref{ablation}, \ref{tab:ablation_detection_perf_mismatch_level}, and \ref{tab:ablation_mismatch_level}). This was due to the 10 patients in the test set who had HCC sizes that were larger than the average size, which facilitated detection by all the model configurations. However, this is not a concerning factor as the test set metrics were consistent for all the models, and the relative scale was maintained throughout the configurations, indicating that none of the models had overfitted to the training set.

\section{Performance on Registered Data}
\begin{figure*}[ht]
\includegraphics[width=\linewidth]{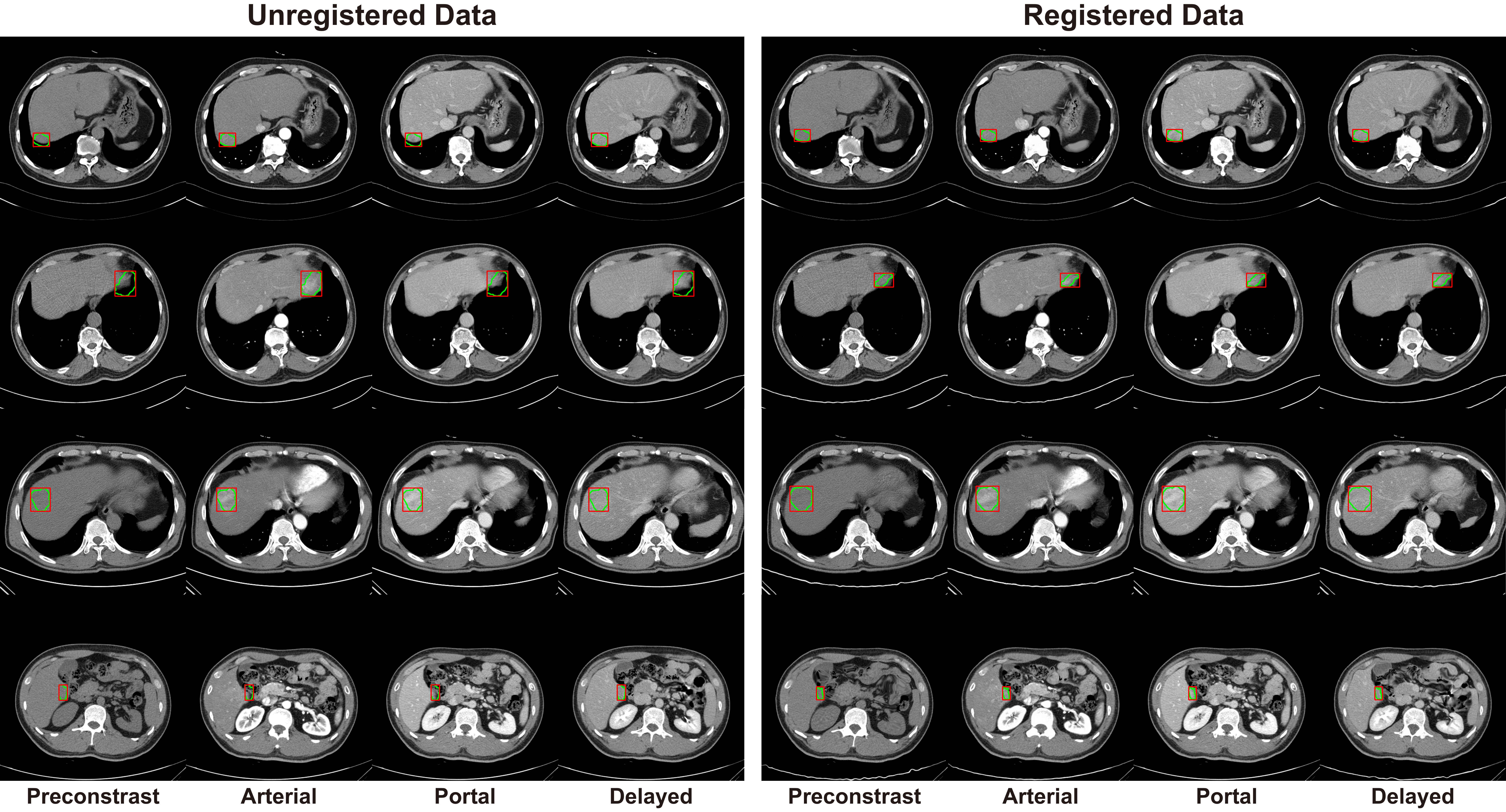}
\caption{
Visualization of registration of multiphase CT dataset. CT images and segmentation masks are registered using the non-rigid method from SimpleElastix with default parameters. Portal phase image is considered as a fixed image. Each row represents a single data point. Green contour corresponds to segmentation masks and red box corresponds to bounding box annotations obtained from the segmentation mask for the object detection task. Left: examples of unregistered images. Right: examples of registered images. After registration, the ground truth labels become consistent with all phases.
}

\label{figure_registereddata}
\end{figure*}

As described in Section~\ref{sec:robust} in main manuscript, we assessed the model’s robustness with a registered variant of our dataset using SimpleElastix~\cite{marstal2016simpleelastix} Python package. Fig.~\ref{figure_registereddata} shows some examples of the unregistered and registered data.
Table~\ref{tab:ablation_detection_perf_mismatch_level} shows the detection performance of the models trained with the registered dataset. GSSD++ and PixelLink++ still retained the state-of-the-art performance on the registered dataset. Table~\ref{tab:ablation_mismatch_level} shows the sensitivity on each registered version of the dataset. The sensitivities of GSSD++ and PixelLink++ were significantly lower than those of the baseline models.

\section{FPN-based Detectors}\label{sec5}
Although the proposed method can be incorporated into other architectures, the architecture based on the FPN \cite{lin2017feature} such as improved RetinaNet~\cite{zlocha2019improving} must be modified significantly, thereby defeating the purpose of comparative evaluation. Specifically, our method performed the best with the deep feature representation at a resolution lower than that of the image space, whereas the FPN featured multiscale object prediction via a lateral connection from a shallow feature map with a significantly higher resolution. In fact, we observed a marginal gain in detection performance when we applied the method to FPN-based detectors. Integrating the proposed method via alternative modules into such an architecture would be an interesting future study.

\begin{table*}[t]
  \centering
  \caption{Detection performance of models trained using the registered version of a multiphase dataset. The performance of each model was evaluated based on the dataset applied to train the model. The dataset was registered using rigid, affine, and non-rigid methods from SimpleElastix. Here, † denotes that segmentation masks were used during training.}
\centering
  \begin{tabular}{l|ccc|ccc|cc}
    \toprule
    \multirow{2}{*}{Methods} & 
    \multicolumn{6}{c|}{Validation Set(\%)} &
    \multicolumn{2}{c}{Test Set(\%)} \\
    & {IoU30} & {IoU50} & {IoU70} & {IoBB30} & {IoBB50} & {IoBB70} & {IoU50} & {IoBB50} \\
    \midrule
    \midrule
    \multicolumn{9}{l}{\textbf{Rigid Method}}\\
    Improved RetinaNet~\cite{zlocha2019improving} + G.Conv~\&~C.Fusion & 60.57 &51.11 &33.52 &63.94 &61.70 &55.00 &62.30 &70.40 \\
    Improved RetinaNet~\cite{zlocha2019improving} + G.Conv~\&~C.Fusion$\dagger$ & 62.63	& 53.53	& 37.77	& 65.24	& 62.99	& 56.27		& 67.56	& 74.10 \\
    GSSD~\cite{lee2018liver} & 58.69 &51.08 &37.15 &60.11 &58.72 &53.44 &64.32 &69.08 \\
    MSCR~\cite{liang2019multi} & 67.14 &56.12 &33.61 &69.46 &60.23 &41.50 &62.02 &65.58 \\
    PixelLink + G.Conv~\&~C.Fusion & 73.60	& 56.39	& 18.92		& 77.28	& 64.77	& 30.13		& 59.07	& 65.12 \\
    GSSD++ (Ours) & 63.27 &53.62 &40.70 &67.71 &65.46 &60.37 &68.14 &77.99 \\
    PixelLink++ (Ours) & 74.85 &57.13 &23.78 &78.27 &64.41 &34.31 &64.58 &75.19 \\
    \midrule
    \multicolumn{9}{l}{\textbf{Affine Method}}\\
    Improved RetinaNet~\cite{zlocha2019improving} + G.Conv~\&~C.Fusion & 62.28 &53.89 &33.69 &65.00 &62.75 &55.41 &61.37 &67.44 \\
    Improved RetinaNet~\cite{zlocha2019improving} + G.Conv~\&~C.Fusion$\dagger$ & 64.30 &54.11 &35.26 &67.42 &65.78 &57.29 &66.81 &75.40 \\
    GSSD~\cite{lee2018liver} & 59.17 &51.17 &35.44 &60.84 &58.84 &53.74 &64.87 &69.33 \\
    MSCR~\cite{liang2019multi} & 70.12	& 53.47	& 20.38		& 74.50	& 60.81	& 30.20		& 60.43	& 64.47 \\
    PixelLink + G.Conv~\&~C.Fusion & 70.12 &51.93 &17.96 &72.31 &58.47 &27.47 &60.00 &65.38 \\
    GSSD++ (Ours) & 63.60	& 55.15	& 40.45		& 66.87	& 64.76	& 59.98		& 68.66	& 76.63 \\
    PixelLink++ (Ours) & 73.10	& 59.83	& 25.20		& 76.21	& 66.87	& 35.33		& 65.43	& 72.30 \\
    \midrule
    \multicolumn{9}{l}{\textbf{Non-rigid Method}}\\
    Improved RetinaNet~\cite{zlocha2019improving} + G.Conv~\&~C.Fusion & 61.29 &55.09 &37.91 &63.92 &62.10 &57.87 &67.25 &70.90 \\
    Improved RetinaNet~\cite{zlocha2019improving} + G.Conv~\&~C.Fusion$\dagger$ & 64.66 &56.71 &38.24 &66.82 &66.78 &61.29 &69.89 &76.75 \\
    GSSD~\cite{lee2018liver} & 63.68 &55.19 &38.08 &66.84 &65.54 &60.24 &67.46 &75.97 \\
    MSCR~\cite{liang2019multi} & 68.82 &59.75 &33.38 &70.96 &63.39 &41.67 &61.56 &66.60 \\
    PixelLink + G.Conv~\&~C.Fusion &72.41 &55.22 &23.12 &75.21 &63.11 &33.65 &59.24 &66.28 \\
    GSSD++ (Ours) & 64.05 &55.63 &41.93 &69.11 &67.33 &62.76 &69.94 &79.34 \\
    PixelLink++ (Ours) & 75.74 &58.46 &24.73 &78.89 &66.13 &35.35 &65.64 &75.10 \\
    
    \bottomrule
  \end{tabular}
  \label{tab:ablation_detection_perf_mismatch_level}%
\end{table*}%

\begin{table*}[ht]
  \centering
  \caption{Assessment of robustness to a registration mismatch. The sensitivity is reported. Here, $\dagger$ denotes that segmentation masks were used during training. A lower performance degradation indicates a higher robustness to a registration mismatch.}
\centering
  \begin{tabular}{l|ccc|ccc|cc|c}
    \toprule
    \multirow{2}{*}{Methods} &
    \multicolumn{6}{c|}{Validation Set(\%)} &
    \multicolumn{2}{c}{Test Set(\%)} & \multirow{2}{*}{Average} \\
    & {IoU30} & {IoU50} & {IoU70} & {IoBB30} & {IoBB50} & {IoBB70} & {IoU50} & {IoBB50} \\
    \midrule
    \midrule
    \multicolumn{10}{l}{\textbf{Rigid Method}}\\
    Improved RetinaNet~\cite{zlocha2019improving} + G.Conv~\&~C.Fusion &4.71 &5.04 &-5.03 &6.03 &5.18 &4.58 &-0.27 &5.82 &3.26 \\
    Improved RetinaNet~\cite{zlocha2019improving} + G.Conv~\&~C.Fusion$\dagger$ &-1.32 &0.78 &5.57 &0.04 &0.23 &-0.66 &6.07 &8.06 &2.35 \\
    GSSD~\cite{lee2018liver} &3.59 &14.41 &2.00 &3.12 &2.62 &0.69 &6.63 &2.54 &4.45 \\
    MSCR~\cite{liang2019multi} &5.09 &8.53 &16.26 &3.39 &4.53 &-0.06 &10.41 &8.49 &7.08 \\
    PixelLink + G.Conv~\&~C.Fusion &5.97	& 9.82	& -5.18		& 6.95	& 13.42	& 1.63		& 2.00	& 1.04	& 4.46 \\
    GSSD++ (Ours) &1.36 &1.34 &1.52 &-0.27 &0.40 &0.63 &0.40 &0.99 &0.80 \\
    PixelLink++ (Ours) &0.98 &-1.12 &-0.96 &0.78 &-1.38 &0.92 &0.60 &2.22 &0.26 \\
    \midrule
    \midrule
    \multicolumn{10}{l}{\textbf{Affine Method}}\\
    Improved RetinaNet~\cite{zlocha2019improving} + G.Conv~\&~C.Fusion &7.33 &9.95 &-4.53 &7.57 &6.75 &5.29 &-1.79 &1.70 &4.03 \\
    Improved RetinaNet~\cite{zlocha2019improving} + G.Conv~\&~C.Fusion$\dagger$ &1.31 &1.85 &-1.14 &3.28 &4.46 &1.13 &5.02 &9.65 &3.19 \\
    GSSD~\cite{lee2018liver} &4.38 &14.56 &-2.74 &4.29 &2.82 &1.25 &7.41 &2.88 &4.36 \\
    MSCR~\cite{liang2019multi} &7.15 &9.88 &17.12 &4.10 &5.57 &0.60 &10.11 &11.25 &8.22 \\
    PixelLink + G.Conv~\&~C.Fusion &1.32 &4.91 &2.36 &3.47 &7.78 &1.86 &4.20 &0.05 &3.24 \\
    GSSD++ (Ours) &1.87 &4.08 &0.90 &-1.52 &-0.69 &-0.02 &1.15 &-0.77 &0.63 \\
    PixelLink++ (Ours) &-1.40 &3.44 &4.72 &-1.90 &2.35 &3.79 &1.90 &-1.69 &1.40 \\
    \midrule
    \midrule
    \multicolumn{10}{l}{\textbf{Non-Rigid Method}}\\
    Improved RetinaNet~\cite{zlocha2019improving} + G.Conv~\&~C.Fusion &5.82 &11.91 &7.12 &6.01 &5.78 &9.31 &7.11 &6.49 &7.44 \\
Improved RetinaNet~\cite{zlocha2019improving} + G.Conv~\&~C.Fusion$\dagger$ &1.86 &6.35 &6.73 &2.41 &5.89 &7.58 &9.20 &11.24 &6.41 \\
GSSD~\cite{lee2018liver} &11.15 &20.78 &4.39 &12.88 &12.76 &11.90 &10.97 &11.37 &12.02 \\
MSCR~\cite{liang2019multi} &7.41 &14.08 &15.68 &5.44 &9.29 &0.36 &9.73 &9.89 &8.98 \\
PixelLink + G.Conv~\ &4.43 &7.91 &13.93 &4.39 &11.14 &11.92 &2.28 &2.78 &7.35 \\
GSSD++ (Ours) &2.56 &4.91 &4.41 &1.77 &3.16 &4.41 &2.96 &2.67 &3.36 \\
PixelLink++ (Ours) &2.14 &1.18 &2.91 &1.56 &1.26 &3.85 &2.21 &2.10 &2.15 \\
    
    \bottomrule
  \end{tabular}
  \label{tab:ablation_mismatch_level}%
\end{table*}%
\end{document}